\documentclass{aastex631}

\shorttitle{Prompt emission and X-ray flares}
\shortauthors{Saji et al.}
\graphicspath{{./}{figures/}}

\begin{document}

\title{Statistical analysis of long GRBs' prompt emission and X-ray flares: multivariate clustering and correlations}

\correspondingauthor{Shabnam Iyyani}
\email{shabnam@iisertvm.ac.in}

\author[0000-0002-8066-8478]{Joseph Saji}
\affiliation{Indian Institute of Science Education and Research, Thiruvananthapuram, Kerala, India, 695551}

\author[0000-0002-2525-3464]{Shabnam Iyyani}
\affiliation{Indian Institute of Science Education and Research, Thiruvananthapuram, Kerala, India, 695551}

\author{Kratika Mazde}
\affiliation{Indian Institute of Science Education and Research, Thiruvananthapuram, Kerala, India, 695551}








\begin{abstract}
The extensive observations done by the X-ray telescope onboard {\it Neil Gehrels Swift} observatory has revealed the presence of late time flares concurrent with the decaying afterglow emission. However, the origin of these flares are elusive. In this work, we made use of the large database of {\it Swift} observations (2005 - 2020) of long GRBs to conduct a systematic statistical study between the prompt gamma ray emission and X-ray flares  by characterising their temporal and spectral properties in terms of duration, quiescent period, peak flux, fluence, minimum variability timescale and spectral power-law index.    
The multi-dimensional database of parameters, thereby, generated was investigated by the principal component analysis which revealed there is no evident correlation between the different parameters of the prompt emission and X-ray flares. 
Furthermore, the correlation studies reveal that while there is a trend of positive correlation
between the minimum variability timescale of flare and its duration, and a strong negative correlation with its peak flux, there are no such correlations observed in the prompt
emission. Similarly, we find a positive correlation between the quiescent period and flare duration, and a negative correlation with the flare peak flux, while no such correlations are observed for the prompt emission of GRBs. Finally, among the X-ray flares, we find two dominant classes whose variations are driven by the minimum variability timescale, peak flux and fluences of the flares. 
A catalog of these different parameters characterising the prompt and flare emissions is presented. 
\end{abstract}



\section{Introduction} \label{sec:intro}

Gamma ray bursts (GRBs) are characterised by two main events: firstly, the prompt emission which composes of the immediate gamma rays and secondly, the afterglow emission which composes of emission ranging from radio till gamma rays \citep{meszaros2006gamma,Kumar_Zhang2015,Iyyani2018}. The quick autonomous re-pointing capability of {\it Neil Gehrels Swift} \citep{Swift_telescope} spacecraft to the location of the burst, in particular, has enhanced the observations linking the prompt and X-ray afterglow emissions of GRBs. This is made possible by the onboard detectors, Burst Alert Telescope (BAT; \citealt{BAT2005} ) and X-ray Telescope (XRT; \citealt{XRT2005}) which make observations in the energy ranges $15 - 150 \, \rm keV$ and $0.3 - 10 \, \rm keV$, respectively.  

One of the intriguing features of the gamma-ray bursts revealed by {\it Neil Gehrels Swift} observations is the presence of flares in the X-ray afterglow lightcurves. 
Systematic survey studies regarding the X-ray flares observed in {\it Swift} XRT were initially carried out by \citealt{First_Swift_flare_catalog} and \citealt{Falcone_etal_2007}. 
X-ray flares are found to occur in both long and short GRBs. This makes the flares a common feature in GRBs despite the difference in progenitors. 
Majority of the flares happen before a few $1000$ s \citep{First_Swift_flare_catalog}, however, in a few of them, they are found to occur as late as $\ge 10^{6}$ s \citep{Swenson_etal_2010,Harsh_etal_2022}. 

Spectral studies such as \citealt{Xrayflares_Morris2008} using wider energy range combining data from BAT, XRT and UVOT\footnote{UltraViolet Optical Telescope (UVOT) onboard {\it Neil Gehrels Swift}.} simultaneously available in a few cases have shown that the flare spectrum requires spectral models more complex than a simple power-law model of the afterglow radiation. This affirmed that the flares are distinctly different from the observed underlying afterglow emission. Several studies have been done to analyse various spectral and temporal properties of these flares observed in X-ray and optical in comparison with the prompt emission properties
\citep{flare_prompt_variability2013,Flare_temp_ppts2015,XrayFlare_study2016,Opticalflares2017,Flare_subclasses2019,multiflare_lag_2021,Spec_flares2022,Statistical_flares2022}. The spectral studies have noted that the flares possess energy fluence comparable to that of the prompt emission \citep{Burrow_etal_2005,Falcone_etal_2007}. In addition, multiple flaring episodes are also found in a certain fraction of GRBs \citep{First_Swift_flare_catalog}. As the flare durations tend to increases with time, the intensity of the flares are found to decrease with time.  

Despite several studies, a self-consistent model explaining the origin, energetics and the evolution of the X-ray flares is yet to be developed. The various proposed models consider the following scenarios: (a) the late time flares are due to delayed central engine activity \citep{MacFadyen_etal_2001,King_etal_2005,Margutti_etal_2011,DallOsso_etal_2017,Gibson_etal_2018}; or if it is an extended phase of the prompt emission episode \citep{Beniamini_Kumar_2016,Mu_etal_2016}; or if it is the late time energy injection into the blast wave by refreshed shocks \citep{Rees&Meszaros1998,Laskar_etal_2015}; or if it is the externals shocks propagating into a dense and uneven circumburst medium \citep{Hascoet_etal_2017,Ayache_etal_2020}.

In this work, we present a detailed multivariate statistical study  of the temporal and spectral properties of the prompt and X-ray flare behaviours. Furthermore, a catalog of the studied properties are presented. Firstly, a reasonably large dataset based on the {\it Swift} observations is selected. Following which the parameters chosen to characterise the prompt and X-ray flare emissions are defined, and the methodology used to extract those features are described in the section \ref{sample_method}. In section \ref{multivariate_analysis}, the multivariate analysis including the distributions of the various parameters, principal component analysis, detailed correlation analysis cum modelling and clustering of X-ray flares are presented. Following, in section \ref{result_discussion}, we discuss and summarise the main results of the analysis and finally, present our conclusions. 


\section{Sample selection and Methodology}
\label{sample_method}

The primary objective of the statistical analysis of 
different parameters of the prompt emission and X-ray flares of the GRBs requires a reasonably large dataset. The {\it Swift} observatory has been discovering 
approximately $100$ bursts per year since 2004 and thereby, the large database including the prompt emission and late time X-ray flares observed by BAT and XRT respectively, 
provides the appropriate database for conducting the statistical studies. The afterglow lightcurves of all the long bursts (lGRBs)\footnote{$T_{90}$ represents the time duration within which $90\%$ of the burst 
emission counts are detected by BAT. Long GRBs are those with $T_{90} \ge 2\, s$.}  detected by {\it Swift} XRT until May 2020 were manually examined 
to look for the presence of flares, which were also confirmed by the flare detection algorithm employed by {\it Swift} XRT team \citep{evans2009methods}. 
After the primary inspection, $340$ lGRBs (about $30\%$ of the total {\it Swift} lGRBs observed during this period) were found to posses 
flaring activity during which the flux evolved in a manner differently from the otherwise continuously decaying afterglow 
behaviour. Among such identified flares, we further sorted and removed certain GRBs with the following light curve behaviours: 
\begin{itemize}
    \item incomplete flaring episodes such as either the rising or decaying phases of the flare light curve is completely absent,
    \item not enough observational data points available in the flare light curve such that it is insufficient for light curve characterisation,
    \item the immediate background emission in the regions pre or post the flare, that is the underlying afterglow emission, is absent or not enough observational data is available,
    \item small peaks in the lightcurve that are of very low signal-to-noise ratio\footnote{Refer section \ref{param} for definition.} ($\ll 0.1$) which are consistent with the statistical variation in the afterglow observations were strictly avoided\footnote{Note: A very small number of bursts (less than 10) with $0.1 <$ SN ratio $< 1$ are included in the sample. In these cases, the parameters were measurable and we have included them in the database, so as to minimise selection biases.}.
\end{itemize}
The above sorting resulted in the removal of about $35\%$ of the initial sample and the final sample consisted of $220$ lGRBs. 
Examples of the afterglow lightcurves of a few excluded GRBs using the above criteria, and selected GRBs with single, double and triple flaring episodes are shown in Appendix A.

\subsection{Parameters of study}
\label{param}

\begin{figure}[h!]
\centering
\includegraphics[width=15cm,height=12cm,keepaspectratio]{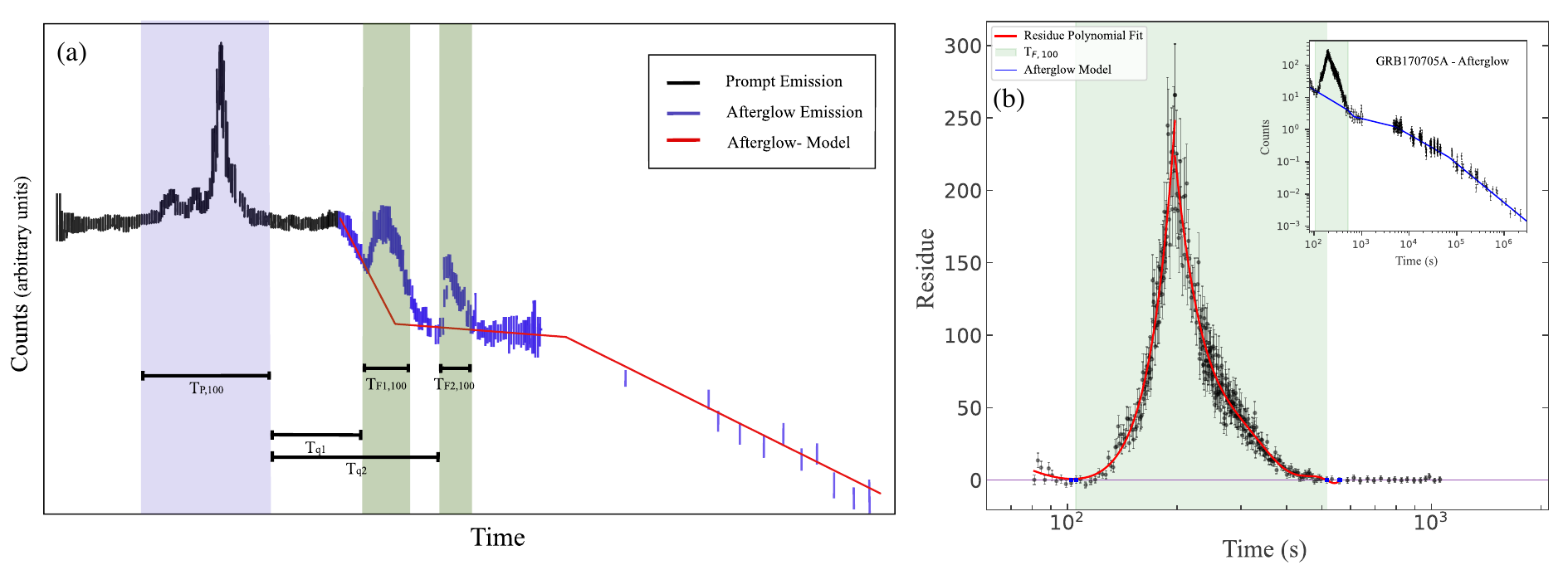}\\
\caption{(a) A schematic representation of the GRB prompt emission and afterglow lightcurves including the late time X-ray flares. 
The temporal parameters, T$_{P,100}$,T$_{F,100}$ and quiescent period [T$_{q}$] are marked. (b) Residue plot of 
GRB170705A showing polynomial fit in the rise and decay phases of X-ray flare is shown. The
points at which the fit coincide with residue = 0 line are marked in blue squares.
The afterglow(black) and its corresponding model(blue solid line) of the same GRB is shown in the embedded plot.}
\label{fig:GRB_schematic}
\end{figure}

A schematic representation of the prompt emission and afterglow ligthcurves consisting of late time X-ray flares is shown in the Figure \ref{fig:GRB_schematic}a.
The following parameters characterising the prompt and X-ray flare emissions of lGRBs are chosen for the study and reported in the catalog:
\begin{itemize}
\item T$_{P,100}$ $\&$ T$_{F,100}$  : The total duration of the gamma-ray burst prompt emission and X-ray flare emission respectively. 
\item Fluence$_{P}$ $\&$ Fluence$_{F}$: The total energy per cm$^2$ obtained after integrating the energy flux over the duration of the burst and flare respectively. 
\item Peak Flux, F$_{P,peak}$ $\&$ F$_{F,peak}$ : The maximum energy flux that is recorded during the burst duration of prompt emission and X-ray flare (subtracting the underlying afterglow flux) respectively. 
\item Peak count time, T$_{P,peak} \, \& \,$ T$_{F,peak}$: It marks the time when maximum count rate is observed in the prompt emission and X-ray flare light curves respectively.
\item  $\alpha_{P}$ $\&$ $\alpha_{F}$: The low energy spectral index of the power-law spectra of the prompt emission and X-ray flare, respectively. 
\item Minimum variability timescale, T$_{P,var}$ $\&$ T$_{F,var}$: It describes the shortest time scale over which a significant change in the count rate of prompt emission and X-ray flare happens, respectively. 

\item Signal-to-Noise ratio of flare: Considering Poisson statistics, the signal-to-noise ratio of the flare is estimated as the ratio of the flare peak count rate (given as the total count rate - afterglow count rate at the peak time) to the square root of 
the total count rate at the peak time. This measure helps to check how distinct and statistically significant the flare is from the underlying afterglow emission. In addition, the distribution of statistical significance allows to verify if the sample is 
biased or not. 

\item Quiescent period, T$_{q}$: It is the difference between the start time of the T$_{F,100}$ of the X-ray flare and the end time of the T$_{P,100}$ of the prompt emission. During this period, there are no source photon counts other than background noises in BAT and the underlying afterglow emission in the XRT. 
\end{itemize}

\subsection{Methodology of Extracting the Parameters}
In the following subsections, the methodology of how the above defined parameters are systematically extracted from the data is presented. 
\subsubsection{ Prompt Emission Data Analysis} 
\label{subsec:Swift-BAT data analysis}                                                           The prompt BAT light curves were analysed using the Heasarc script of Bayesian Block binning named as {\fontfamily{qcr}\selectfont battblocks} with {\fontfamily{qcr}\selectfont ncp$\_$prior = 6} and uniform time binning of $0.01$s to estimate the total duration, $T_{P,100}$, and the peak time, $T_{P,peak}$ of the prompt emission of the burst.                                        
The time integrated and peak count BAT spectra of the burst were created for the duration of $T_{P,100}$ and for 1 second duration around $T_{P,peak}$ of the prompt emission in the energy range $15-150$ keV, respectively. The spectral data were then analysed using the Multi-Mission Maximum Likelihood Framework (3ML, V.2.2.1, \citealt{Vianello_etal_2015}) software package. 
Using 3ML, the spectra were analysed using different phenomenological spectral functions such as simple power-law, cutoff power-law and Band function \citep{Poolakkil_etal_2021}.
By comparing the Akaike Information Criterion (AIC)\footnote{AIC is the statistic estimator which is calculated using the formula: $2k-2 ln(\hat{L}) $ where $k$ is number of estimated parameters and $\hat{L}$ is the maximum likelihood function.}, the spectral function fit with the minimum AIC was considered as the best fit spectral model. 
From the best fit model, the low energy power-law index, $\alpha$ of the time integrated spectrum and the total energy fluence were noted. The peak energy flux was estimated from the analysis of the peak spectrum generated around the peak time of the prompt emission. 

\subsubsection{ X-ray Flare Data Analysis} \label{subsec:Swift-XRT data analysis}

The pre-processed XRT count rate light curves for the afterglow analysis were obtained from {\it Swift} XRT repository\footnote{Data products are 
available in the link: https://www.swift.ac.uk/xrt$\_$live$\_$cat/} in the energy range, $0.3-10$ keV. The binned GRB XRT afterglow light curves are ensured to contain at least 15 counts and the errors are calculated using the Gaussian statistics \citep{evans2009methods}. The temporal structure of the 
afterglow emission consists of several power-laws of decaying count rate which is entirely different from the highly varying erratic 
pulses of emission in the flares (Figure \ref{fig:GRB_schematic}a). Therefore, the first step in processing the data was to separate the 
two emissions. This was achieved by removing the rough data region that included the flares from the entire afterglow light curve and 
modelling the remaining smoothly decaying afterglow emission using a power-law or a combination of power-laws 
\citep{evans2007online,evans2009methods}. The best-fit models were determined using the 3ML 
and AIC statistics. A multi-broken power-law model was found to best 
represent the underlying afterglow emission. The resulting best-fit model of the afterglow light curve was subtracted from the total 
afterglow emission, which produced a residual light curve including the propagated errors  (Figure \ref{fig:GRB_schematic}b). The peak count time of the flare was used to identify the rise and decay 
phase of the flare. Using the
technique of fitting a polynomial function to the residual counts light curve, the respective last and first intersection points with the background curve in the rise and decay phases are 
selected as the start and the stop times of the flare episode respectively.  The best fit polynomial for the rise and decay phases of the flare was selected after an iteration of 
polynomial fits of different orders and the one which gave the reduced chi-square 
closest to 1.

To ascertain the variation in the estimated start and stop times of the flare episode, 5000 MonteCarlo simulated 
residual light curves were generated by assuming a Gaussian distribution\footnote{ The assumption of Gaussian distribution is applied to each data point of count rate in the residual light curve, by taking into account that if in case the count rate takes a negative value during 
the simulation, it would still have a physical meaning implying that the afterglow emission is more dominant at that instance. In addition, count rate allows for fractional values.} to each data point where the original residual 
count is the mean and its associated error as the standard deviation. 
The rise and decay phases in each residual light curve were analyzed using polynomial fits with the same best-fit order as the original residual light curve.
The mean and standard deviation of the distributions of the start and stop times of the flare episode is referred to as T$_{F,start}$ and T$_{F,stop}$, and their associated errors respectively. 
The distribution of the duration of the flare episode was generated by taking the difference between the start and stop times of the flares of each simulated residual count curve. 
The mean and standard deviation of the distribution of the duration of the flare is reported as the T$_{F,100}$ and its associated error.

The X-ray flare spectrum was generated for the whole duration, T$_{F,100}$ of the flare in the energy range $0.3-10$ keV using the {\it Swift}-XRT 
repository and analysed in 3ML. In order to separate the underlying afterglow emission, the X-ray spectra was firstly 
generated in regions selected pre and post of the X-ray flare or either one of them, depending on if enough data were available. 
The underlying afterglow emission spectra was first analysed 
using a power-law function multiplied by two absorption terms: galactic, $n_{H}$ and source, $n_{H}$. The galactic, $n_{H}$, was chosen based on the right-ascension (RA) and declination (Dec) of each burst whereas, the source, $n_{H}$ was left as a free 
parameter in the fit. The  Galactic $n_{H}$ values were obtained from the Galactic $n_{H}$ calculator available at {\it Swift} science data centre\footnote{https://www.swift.ac.uk/analysis/nhtot/}, which is based on \cite{Willingale2013}.
Thereafter, the X-ray spectra generated for the flare duration is analysed using two power-law functions multiplied by the absorption coefficients (source, $n_{H}$ was fixed at the value obtained during the afteglow spectral fit). One of the power-law functions was frozen to the afterglow spectral fit values and the other power-law function was left free to model the flare spectrum. Thereby, using the flare power-law model, the respective energy flux and fluence in the energy range, 
$0.3-10$ keV, were estimated. 
In addition, the spectral index, $\alpha_F$, along with the above mentioned parameters were added to the database. The peak flux of flare corresponding to the peak time 
is calculated from the flux lightcurve obtained from the {\it Swift}-XRT repository after subtracting the average afterglow flux 
\label{subsec:Late time prompt emission and Multiple Flares}

In the current sample, there are GRBs with multiple episodes of X-ray flare emissions. The end of a flaring episode is defined when the decay phase of the light curve touches the background. The further rise after this instance or after certain period of quiescence is considered as the onset of the consecutive flaring episode (Figure \ref{fig:GRB_schematic}a). Using this criteria, we identify multiple episodes of the flare. 
Out of the $220$ GRBs in the sample, there are $44$ GRBs which possess two flaring episodes  and $4$ GRBs with three flaring episodes.
In $96$ GRBs, late time X-ray flares were observed concurrently by the BAT and XRT detectors. In such cases, the emission observed in BAT exempting the X-ray flare detected in XRT, is considered as the prompt emission in our study. 
Therefore, the $T_{P,100}$ estimate in such cases considers only the burst duration of the first episode of radiation that is observed in BAT. The quiescent period, $T_q$ for the various flaring episodes are respectively measured as the difference between the onset time of the flaring episode and the end time of the $T_{P,100}$ of the burst (Figure \ref{fig:GRB_schematic}a). 

\subsubsection{Variability Measurement} \label{Variability Maasurement}
One of the defining characteristics of gamma ray burst emission is its highly variable light curves observed during the 
prompt emission \citep{MacLachlan_etal_2013}. Interestingly, the X-ray flares also exhibit strong variability and erratic light curves \citep{Sonbas_etal_2013}. The minimum 
timescale of variability is the minimum timescale over which there is a significant or reasonable change in count rate in the observed light curve. The study of the variability pattern can provide us the information regarding the size and 
location of the source region \citep{Kobayashi_etal_1997}.

In this work, the Bayesian block (BB) technique \citep{Scargle2013} of binning the counts light curve to identify the minimum variability timescale of both the prompt as well as the X-ray flare emissions was employed. BB bins the light curve such that 
each time interval possess a constant Poisson rate. This methodology divides the light curve into various time intervals of dynamical time widths and signal-to-noise ratio and closely follows the true underlying variation in the emission \citep{Burgess2014,Vianello_etal_2018}. 
The time interval with the minimum timescale is considered as the minimum variability ($T_{var}$) of the light curve. 
The light curves of prompt and flare emissions are analysed using the Bayesian Block binning method available in Astropy package \citep{2022ApJ...935..167A} in python.
In order to compare the estimated variability timescales of both the prompt and flares emissions of a GRB as well as between 
different GRBs in the sample, a common false alarm probability of $p_0 = 0.01$ is adopted throughout the analysis. The $p_0 = 0.01$ is 
found to be an optimal choice allowing to effectively capture the variabilities across different emissions and provides a confidence level of $99\%$. This approach, thereby, ensures a 
systematic analysis, avoiding the use of multiple configurations specific to different GRBs.


\section{Multivariate Analysis}
\label{multivariate_analysis}
In this section, we analyse the multi-dimensional database that is generated via the methodology described in the previous section to assess the probability distributions, correlation and clustering of the various studied parameters. 

\subsection{Distributions of prompt and X-ray flare parameters}
\begin{figure}[h!]
\centering
\includegraphics[width=15cm,height=6cm]{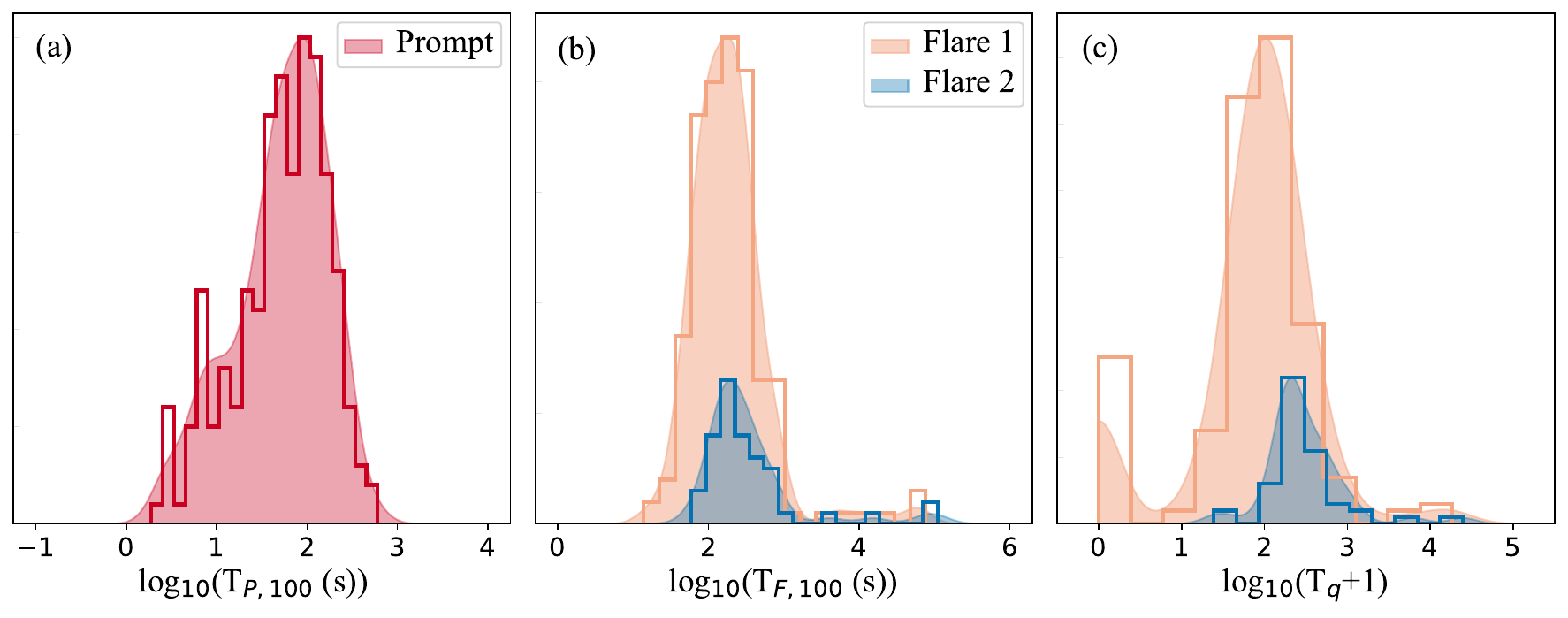}\\
\includegraphics[width=10cm,height=6cm]{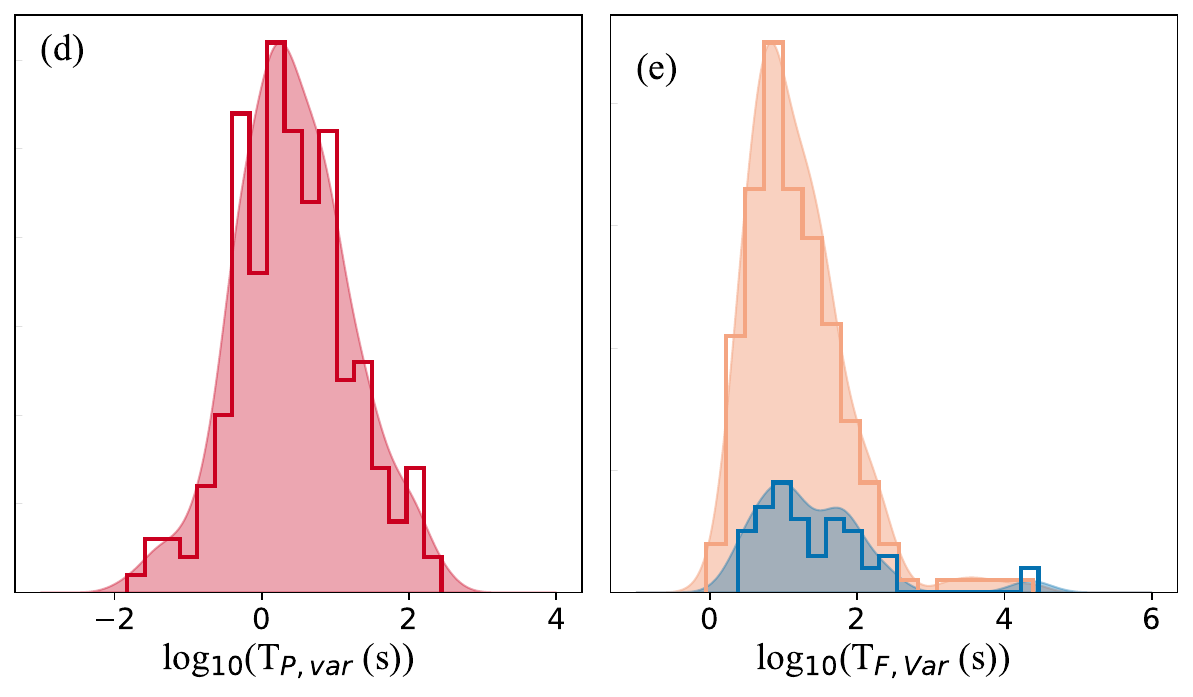}\\
\includegraphics[width=15cm,height=11cm]{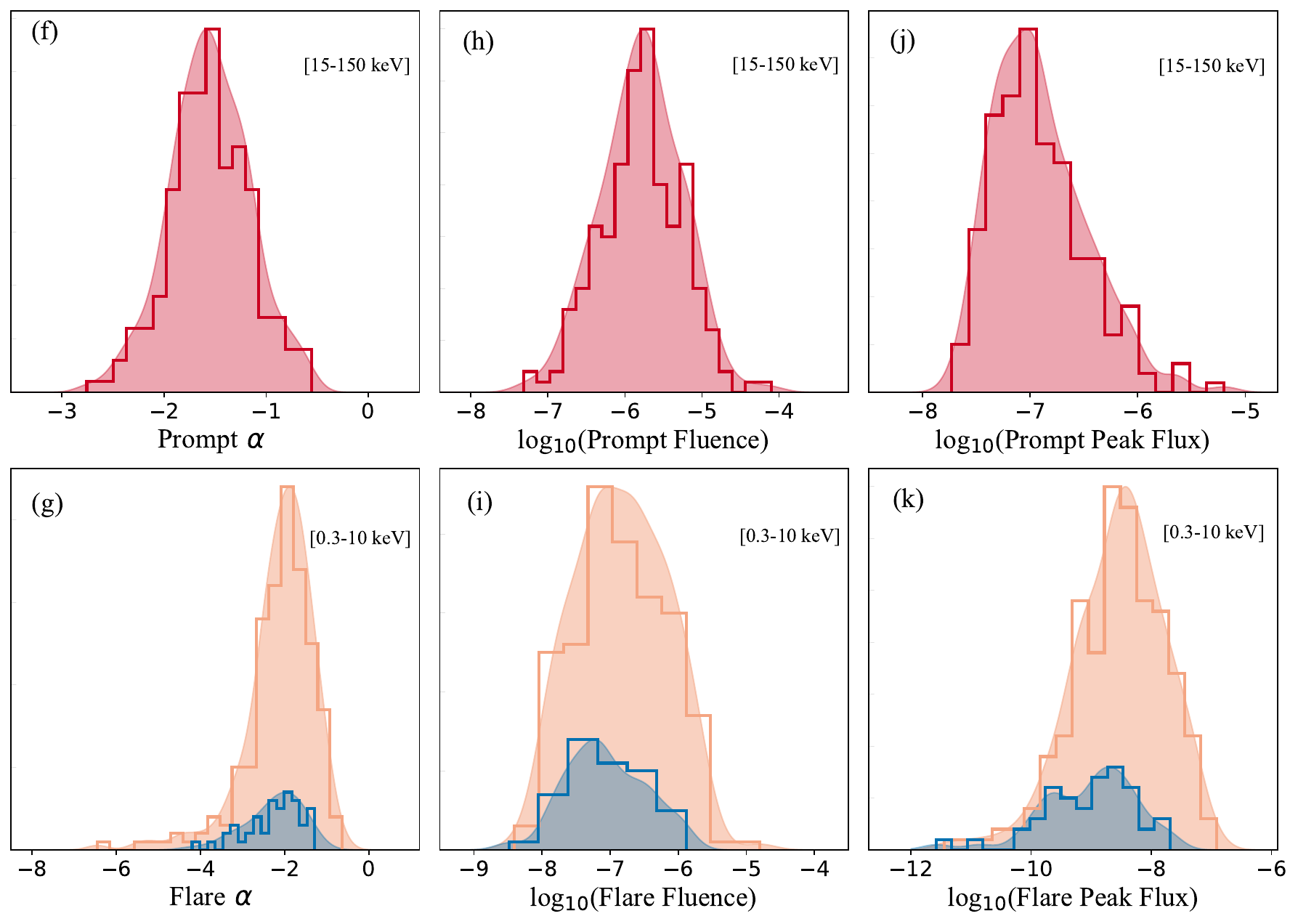}\\

\caption{The distributions obtained for the prompt (red) and flare (orange for first flare episode and blue for second flare episode) temporal parameters are shown. 
The vertical bars represent the histograms and the respective KDE curves are shown in color filled shaded curves. The distributions of 
a) T$_{P,100}$, b) T$_{F,100}$, c) T$_{q}$, 
d) T$_{P,var}$, e) T$_{F,var}$, f) $\alpha_{P}$,  
g) $\alpha_{F}$, h) $Fluence_P$, i)   $Fluence_F$, j) $F_{P,peak}$, and k) $F_{F,peak}$ are shown.}
\label{fig:dist}
\end{figure}

The histogrammed probability density plots of the temporal and spectral related parameters of both the prompt and X-ray flare emissions studied for the GRBs in this work are presented in the Figure \ref{fig:dist}. In addition, the smoothed version of the distribution obtained via the kernel density estimation (KDE) using the Gaussian kernels are also shown. 


The X-ray flare episodes have durations on average greater than that of the prompt emission and the consecutive flaring episodes tend to be even longer. The mean (median) of 
the prompt duration is around $51^{+117}_{-35}  \, \rm s$ ($59 \, \rm s$), whereas the first and second X-ray flaring episodes, on average (median), have longer durations of $188^{+449}_{-133} \, \rm  s$ ($169 \, \rm s$) 
and $344^{+1182}_{-267} \, \rm s$ ($230 \, \rm s$), respectively (Figure \ref{fig:dist}a and  \ref{fig:dist}b). The average (median) quiescent periods of the first and second flaring episodes are found to be around $ 69^{+368}_{-58}  \, \rm s$ ($91 \, \rm s$)  and $ 303^{+593}_{-200}  \, \rm s$ ($224 \, \rm s$), respectively (Figure \ref{fig:dist}c).
One of the immediate similarities that can be 
noted between the prompt emission and the X-ray flares via a visual inspection is the erratic nature of both the light curves which is quantified using variability timescale measurements. The prompt emission 
light curves exhibit minimum variabilitPrompt emission and X-ray flaresies over time scales on average (median) $ 2.7^{+13.4}_{-2.2}  \, \rm s$ ($2.4 \, \rm s$), see Figure \ref{fig:dist}d, while the minimum variability timescales 
of first and second X-ray flare episodes are relatively of longer durations on average (median)  $ 15^{+56}_{-12}  \, \rm s$ ($11 \, \rm s$) and $ 25^{+144}_{-22}  \, \rm s$ ($16 \, \rm s$), respectively (Figure \ref{fig:dist}e). We note that X-flare light 
curves depending on the exposure times, for majority of first episodes an average $1\, s$ time resolutions were available, but, in some cases, the time resolutions were longer.

The low energy power-law spectral indices, $\alpha_P$ for the time integrated spectrum of the total prompt emission duration (energy band: 15 - 150 keV) are found to be on average (median) around $-1.5 \pm 0.40$ (-1.5), see Figure \ref{fig:dist}f. However, the 
lower energy power-law spectral indices, $\alpha_{F1}$ and $\alpha_{F2}$ of the time integrated spectrum of the total first and second X-ray flare episodes in energy range 0.3 - 10 keV are on average (median) much softer, around $-2.1 \pm 0.83$ ($-2.0$) and $-2.25 \pm 0.65$ 
($-2.17$), respectively (Figure \ref{fig:dist}g). 

The fluence and the peak flux of the prompt episode is on average (median) around $1.6^{+3.8}_{-1.1} \times 10^{-6}$ ($1.7 \times 10^{-6}$) $\rm erg/cm^2$ and $1.2^{+2.1}_{-0.78} \times 10^{-7}$ ($1.1 \times 10^{-7}$) $\rm erg/cm^2/s$, respectively in the 
energy range $15$ - $150$ keV (Figures \ref{fig:dist}h and \ref{fig:dist}j). 
The average (median) of the fluence of the first and second X-ray flaring episodes in 0.3 - 10 keV is around 
 $1.5^{+5.2}_{-1.1} \times 10^{-7}$ ($1.5 \times 10^{-7}$) $\rm erg/cm^2/s$ and $8.7^{+24.7}_{-6.5} \times 10^{-8}$ ($7.2 \times 10^{-8}$) $\rm erg/cm^2$, respectively (Figure \ref{fig:dist}i). 
The first and second episodes of the X-ray flare episodes have peak flux on average (median) around 
 $2.9^{+14}_{-2.4} \times 10^{-9}$ ($3.2 \times 10^{-9}$) $\rm erg/cm^2/s$ and $9.3^{+47}_{-7.8} \times 10^{-10}$ ($1.2 \times 10^{-9}$) $\rm erg/cm^2$, respectively (Figure \ref{fig:dist}k).

\subsection{Principal component analysis}
\label{PCA1}
\begin{figure}[h!]
\centering
\includegraphics[width=15cm,height=10cm,keepaspectratio]{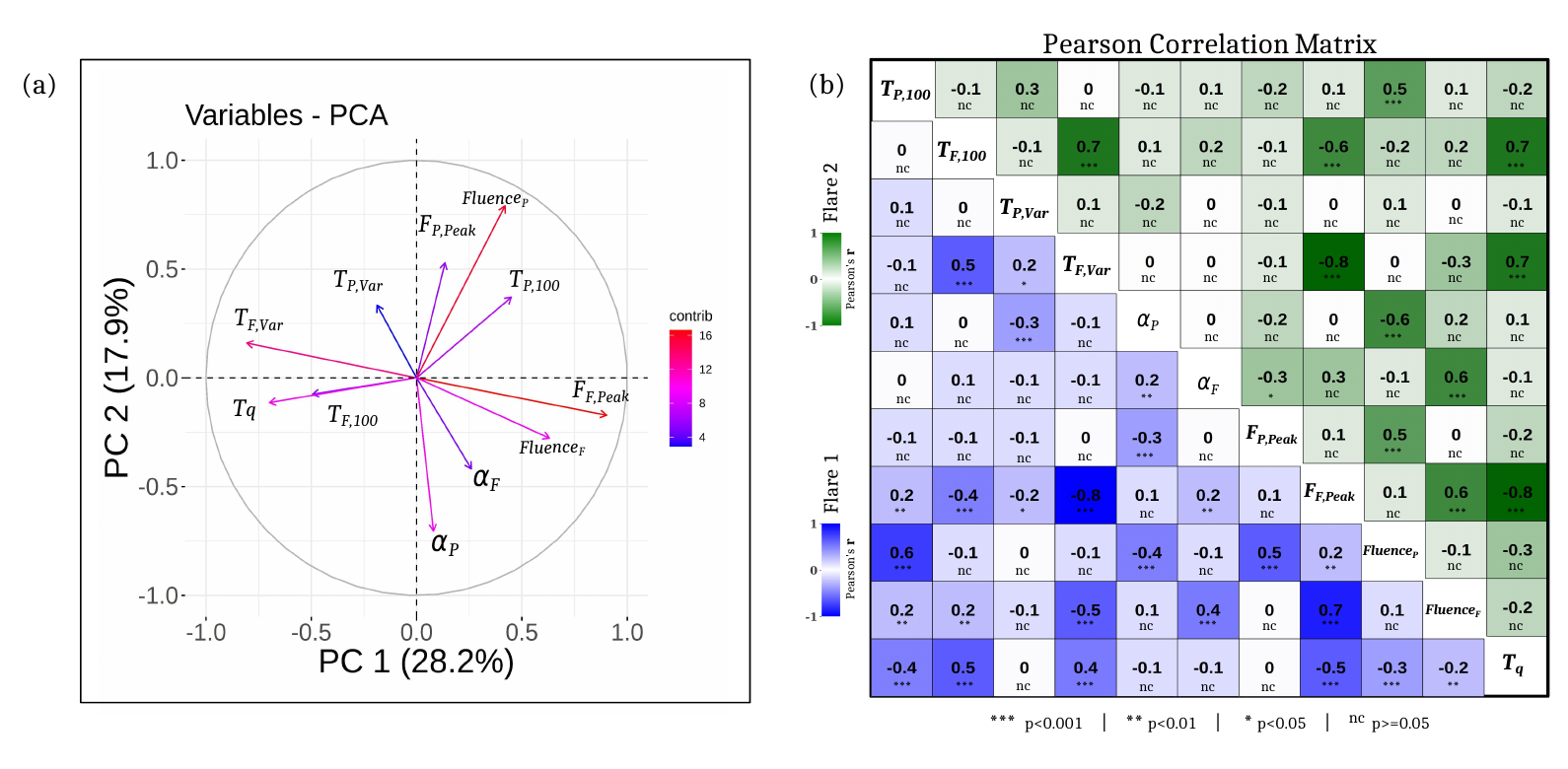}\\
\caption{(a) PCA circle of correlation showing the relationship between all the prompt and X-ray flare parameters is shown. 
The percentage of the total contribution of each parameter to the PCA axes is represented by the color coding. (b) The Pearson Correlation Matrix of the various parameters of the prompt emission and the first (blue), and second (green) X-ray flare episodes are shown. Markers representing the p-values of the obtained correlations are shown in each cell representing the statistical significance of the correlation.}
\label{fig:PCA_full}
\end{figure}

Principle Component Analysis (PCA) allows to search for trends or correlations between parameters and sort them based on their contribution to the observed variance in the data. 
The PCA of the entire multi-dimensional dataset including the parameters of both prompt and X-ray flares, thereby, gives us some insight into the relation between different 
parameters and also shows the parameters that mainly drive the variability in clustering. PCA reduces 
the multi-dimensionality of the dataset into a few principal components which is a linear combination of the original parameters. We find that principal component 1 (PC1) and principal component 2 (PC2) 
together account for $46\%$ of the total variability in the dataset. Using PC1 and PC2, the PCA circle of correlation is created as shown in Figure \ref{fig:PCA_full}a. Here the correlation refers to the 
Pearson correlation coefficient 
between the parameter and the principal components. Thereby, the parameters that are positively 
correlated among each other are grouped together within the same quadrant and if they are negatively correlated the parameters would be positioned in opposite 
quadrants. The parameters that are less correlated with each other would be positioned in adjacent quadrants and would be at nearly right angles to each other. It is also key to note that the length of 
each vector variable from the origin (centre of the circle) in the plot represents the quality of 
representation of the variable. In other words, it represents the percentage of the contribution of that 
parameter in defining the PCA axes 1 and 2 combined, which is provided in the contribution color bar shown in the Figure \ref{fig:PCA_full}a.  

With the above understanding, from the PCA analysis\footnote{For the PC analysis, the R package {\fontfamily{qcr}\selectfont FACTOMINER} \citep{factominer} was used.} of the dataset, we find there exists no explicit correlation between the temporal and spectral parameters of prompt emission and X-ray flares, as evident from Figure \ref{fig:PCA_full}a . 
In case of $\alpha$ and $T_{var}$, it is important to note that the quality of representation of these parameters on PC1 and PC2 are lowest and therefore, their degree of correlations may not be strong (see section \ref{correlations} for more details). Using this initial results of PC analysis as a guide, we look into the correlations between the various parameters and clustering in the subsequent sections in detail.    

\subsection{Correlation Analysis}
\label{correlations}

The correlation analysis explores and identifies trends, patterns or relationships between different parameters under consideration, including positive, negative or no correlation. This can be quickly assessed by visualizing data in a scatter plot. Most importantly, correlation does not imply direct causation, as the relationship between parameters may be complex. Nonetheless, a correlation does indicate interdependence, allowing for further exploration of the underlying physical scenario.

A Pearson Correlation Matrix of the various parameters of the prompt emission and the first and second X-ray flare episode is shown in Figure \ref{fig:PCA_full}b. The correlation matrix features markers indicating the various levels of statistical significance, including p-value $<0.001$, $p<0.01$, $p<0.05$, and $p >0.05$. The p-value represents the probability of the null hypothesis which considers there is no correlation between the variables and thereby serves as the measure of the statistical significance of accepting the alternate model i.e there exists a correlation between the two variables. The parameters of the third flare episodes are not included in the 
correlation matrix because only a small fraction of the sample consisted of such multiple flaring episodes. However, they are included in the detailed correlation plots and modelling. In this work, we have modelled only those correlations between parameters that exhibited a positive or negative Pearson correlation coefficient $> 0.5$, both in first and second flaring episodes. All the modelled correlations in our investigation demonstrate strong statistical significance as indicated by the low p-values, less than 0.001.  

\begin{figure}[h!]
\centering
\includegraphics[width=16cm,height=20cm,keepaspectratio]{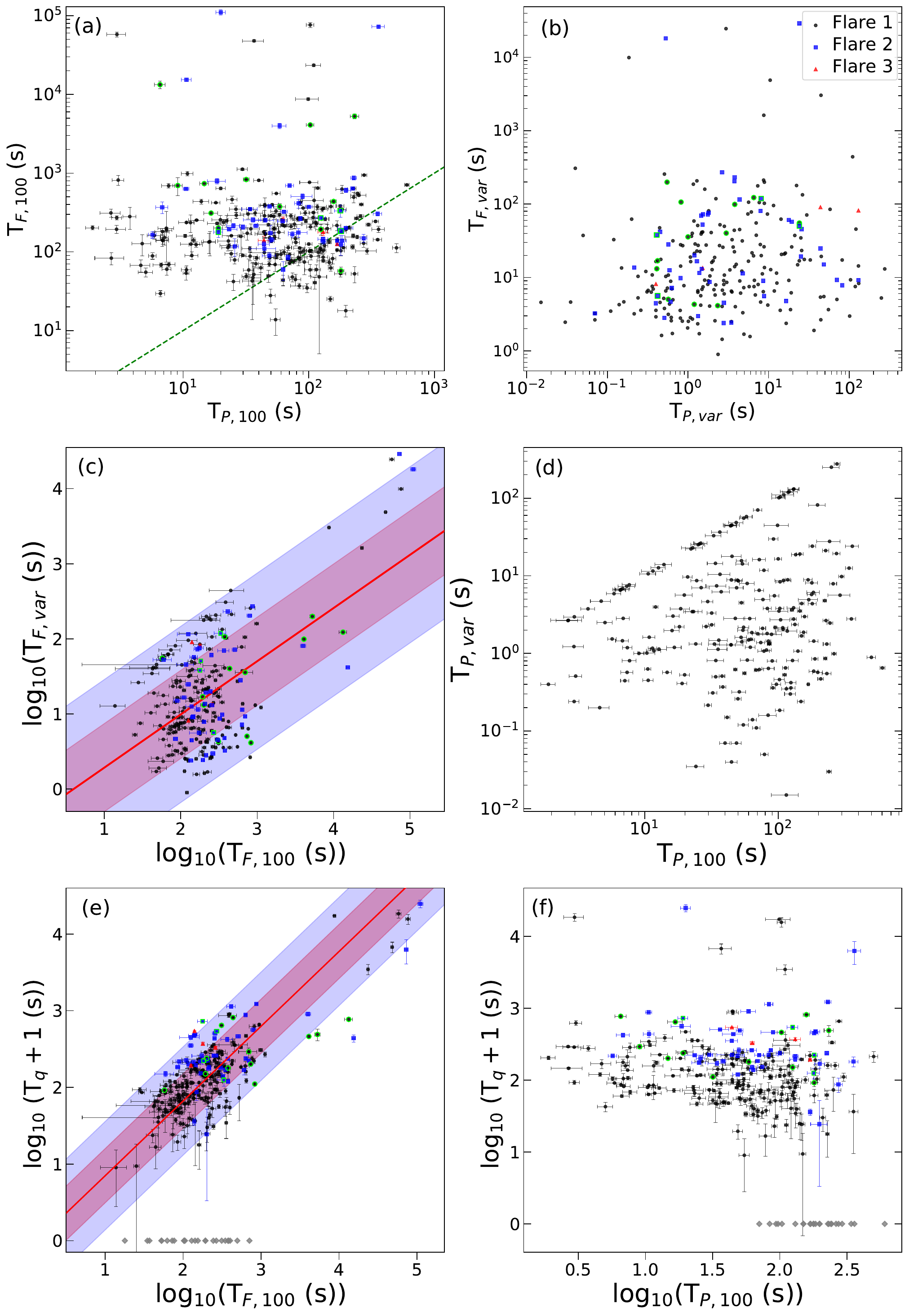}\\
\caption{ (a) T$_{F,100}$ is plotted against $T_{P,100}$. The black circles, blue squares and red triangles represent the first, second and third X-ray flaring episodes, respectively. The same color code and markers are used hereafter.The green dashed line represents when T$_{P,100}$ = T$_{F,100}$. (b) The $T_{F,var}$ is plotted against the corresponding $T_{P,var}$.   
(c)The $T_{F,var}$ is plotted against $T_{F,100}$. 
The linear fit in the log space is shown by red solid line. The shaded red and blue regions correspond to 1$\sigma$ and 2$\sigma$ deviations of data from the fit. The color code represents the same contours hereafter.  d)  T$_{P,100}$ is plotted against $T_{P,var}$
(e) T$_{q}$ is plotted against T$_{F,100}$ in the left hand panel and (f) against T$_{P,100}$ in the right hand panel respectively. 
Note that the grey diamonds represent the GRBs with quiescent period equal to zero, are not included in the modelling of the correlation. The parameters of flares that occur during the plateau phase of the afterglow is marked in green unfilled circle and square markers for the first and second episodes of flare respectively}
\label{fig:Dur_multi}
\end{figure}

 
In Figure \ref{fig:Dur_multi}a, the duration of the X-ray flare, $T_{F,100}$, is plotted against the duration of the respective prompt emission, $T_{P,100}$, of the GRB. No particular correlation can be observed. For the purpose 
of comparison, the identity line representing when $T_{P,100} = T_{F,100}$ is marked. The 
duration of the flares including both first, second and third episodes of emission largely lie above this identity 
line suggesting that the flaring episodes tend to have durations greater than their respective prompt emissions. In addition, in case of multiple episodes of flaring, the consecutive episode is always longer than the preceding episode of flaring.


 As evident from the Figure \ref{fig:Dur_multi}b, the flare minimum variability does not show 
any explicit correlation between the temporal properties of the prompt emission properties such as its minimum variability and duration (see Figure \ref{fig:PCA_full}b). In comparison to the quiescent period, there is only a trend of positive correlation between $T_{F,var}$ and $T_q$, with a weak Pearson correlation coefficient of $+0.4$ (Figure \ref{fig:PCA_full}b). 
The minimum variability timescale of the flares is compared with the total duration of the flare as shown in Figure \ref{fig:Dur_multi}c. We note there is a positive correlation between $T_{F,var}$ and $T_{F,100}$ as follows 
\begin{equation}
   log_{10}T_{F,var} = (+0.71 \pm 0.06)\, log_{10}(T_{F,100}) - 0.42 \pm 0.15 
\end{equation}
\noindent
We note that this positive correlation is a weak trend as we are 
limited by the small number of flares with very long durations ($> 1000\, \rm s$). Thus, this correlation can be affirmed in the future with increased detections of flares with longer durations and lightcurves with smaller time resolutions. Interestingly, in 
contrast to this behaviour of flares, we note that there exists no significant correlation between the prompt emission minimum variability and its duration (Figure \ref{fig:Dur_multi}d) as 
evident from the Pearson correlation matrix shown in Figure \ref{fig:PCA_full}b. For a small number of GRBs, the BB binning of prompt emission resulted in minimum timescale equivalent to the whole burst duration (Figure \ref{fig:Dur_multi}d). 

A positive correlation is found between the quiescent period,  $T_q$,  and the X-ray 
flare duration, $T_{F,100}$ (Figure \ref{fig:Dur_multi}e). The positive correlation is modelled using a linear fit in the log-space as follows
\begin{equation}
   log_{10}(T_q+1) = (+0.97 \pm 0.05)\, log_{10}(T_{F,100}) - 0.12 \pm 0.10 
\end{equation}
\noindent
and is shown in solid red line in Figure \ref{fig:Dur_multi}e. 
In comparison, there is only a weak trend of negative correlation observed between the quiescent period and the prompt emission duration of the bursts as shown in Figure \ref{fig:Dur_multi}f.

The prompt and flare emission spectra were modelled using the BAT and XRT data, 
obtained in the energy ranges 15-150 keV and 0.3-10 keV respectively. For the purpose of comparison of spectral behaviours, we assumed that the obtained prompt spectral model 
remains the same in the extrapolated lower energy range of 0.3 - 10 keV and thereby, studied the correlations between the spectral properties of the prompt and flare emissions. 

\begin{figure}[h!]
\centering
\includegraphics[width=.9\textwidth,height=16cm,keepaspectratio]{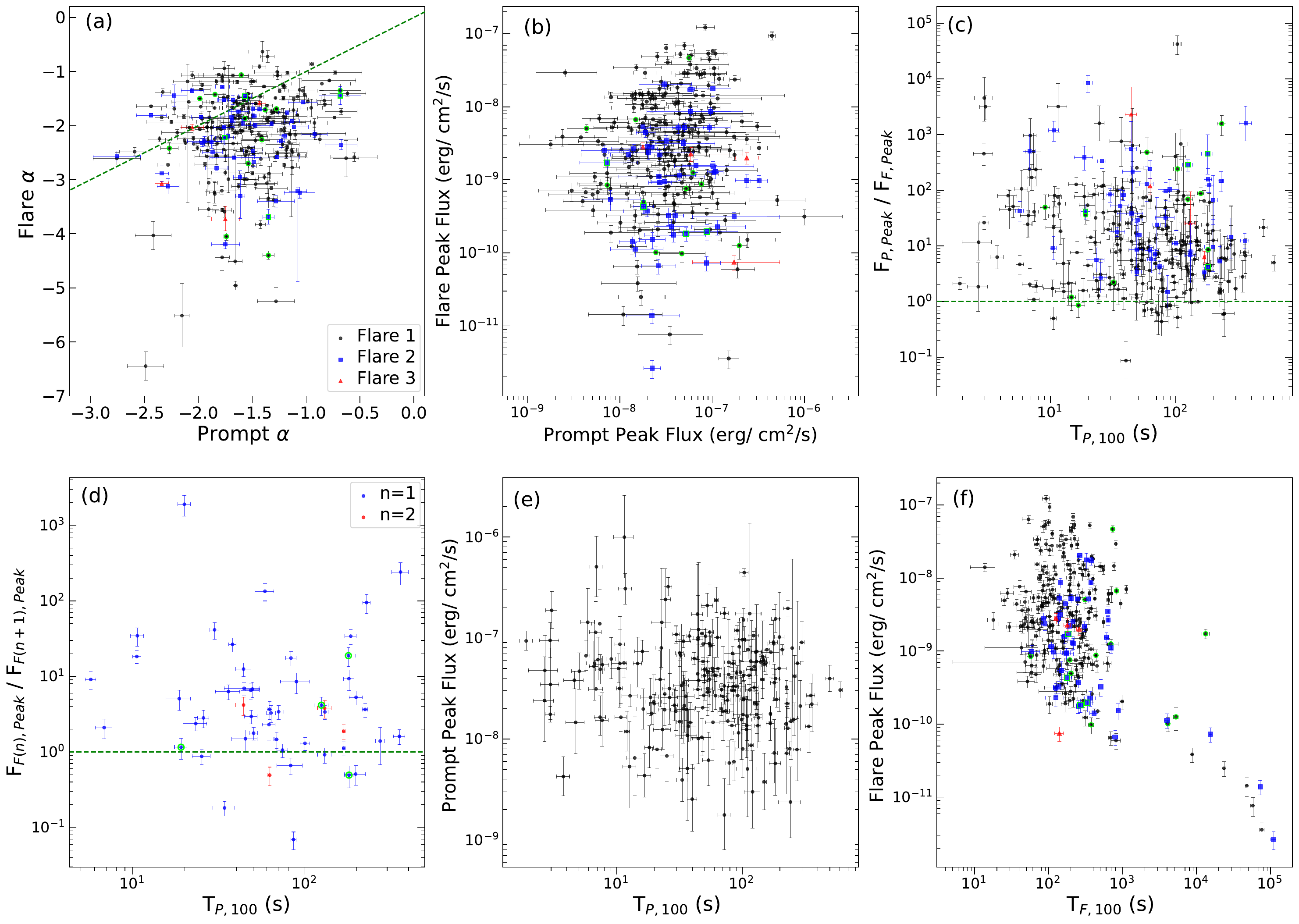}\\
\includegraphics[width=.6\textwidth,height=16cm,keepaspectratio]{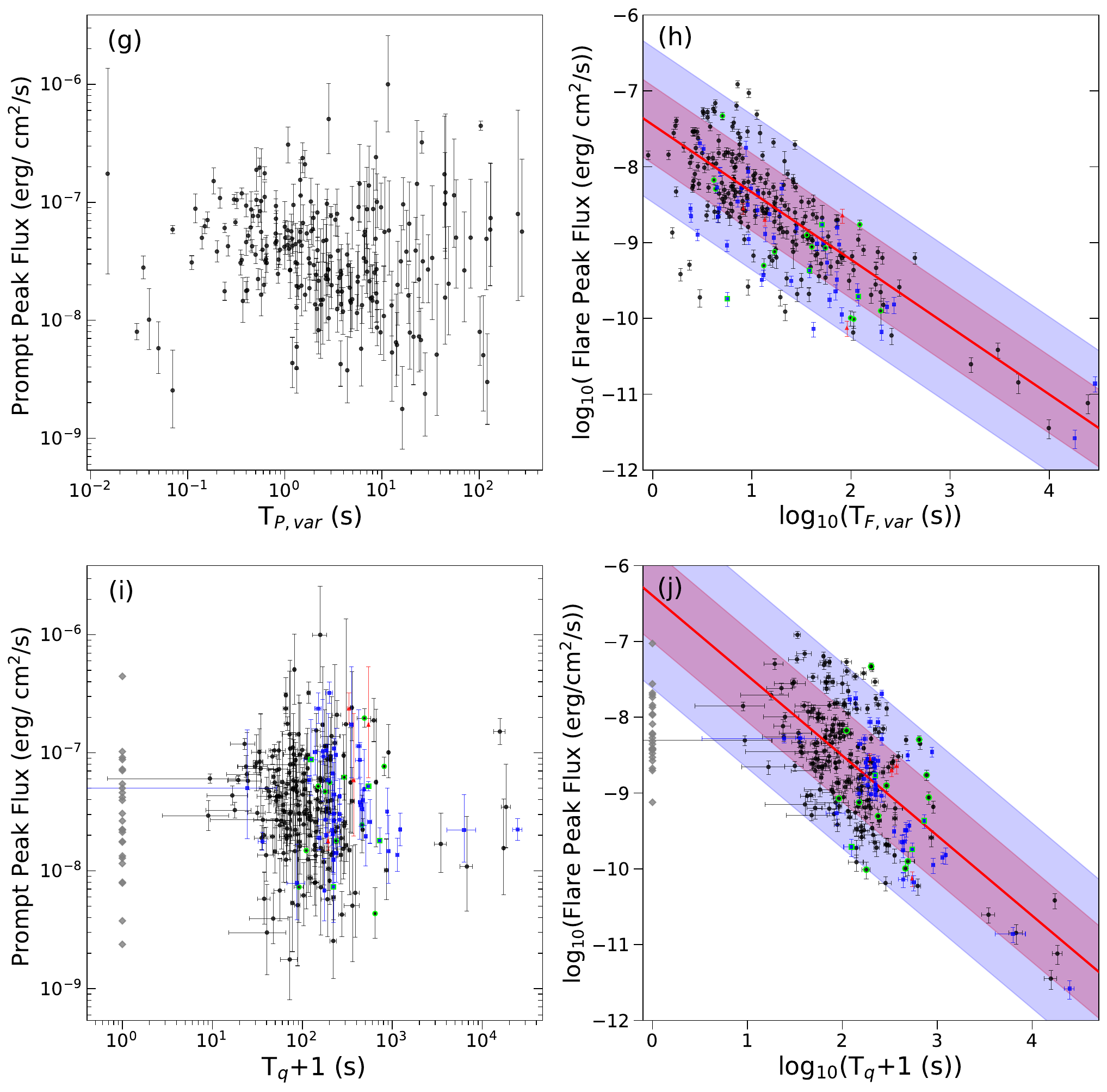}\\
\caption{(a) $\alpha_{P}$ is plotted against $\alpha_{F}$. (b) F$_{P,peak}$ is plotted against F$_{F,peak}$. (c) The ratio of F$_{P,peak}$ to F$_{F1,peak}$ and F$_{F2,peak}$ is plotted against the T$_{P,100}$ in black circles and blue squares respectively. (d) The ratio of F$_{F1,peak}$ to F$_{F2,peak}$ is plotted against T$_{P,100}$ in blue circles and the ratio of F$_{F2,peak}$ to F$_{F3,peak}$ in red circles. (e) F$_{P,peak}$ is plotted against T$_{P,100}$. (f) F$_{F,peak}$ is compared against T$_{F,100}$. (g) F$_{P,peak}$ is plotted against T$_{P,var}$. (h) F$_{F,peak}$ is compared against T$_{F,var}$. (i) F$_{P,peak}$ is plotted against T$_{q}$. (j) F$_{F,peak}$ is compared against T$_{q}$. The linear fit modelling the correlations in (h) and (j) are shown in red solid lines. The parameters of flares occurring during the plateau phase of the afterglow are highlighted in green unfilled circle and square markers for the first and second episodes of flare respectively.}
\label{fig:Energy_corr_multi}
\end{figure}

Firstly, we studied the correlation between the low energy power-law index ($\alpha_P$) of either the power-law or the cutoff power-law model used to best model the time integrated prompt spectra in the energy range $15 - 150$ keV and the low energy power-law index 
($\alpha_F$) of the power-law model that was used to model the time integrated flare spectra in the energy range $0.3 - 10 $ keV, as shown in the Figure \ref{fig:Energy_corr_multi}a. There is no explicit correlation observed between the $\alpha_F$ and $\alpha_P$. For majority of the sample the $\alpha_F$ tends to be much softer than $\alpha_P$.  


The high energy ($> 10 keV$) part of the flare spectrum is not available for analysis and therefore, the peak of the flare spectrum is not known. It has been observed in the prompt emission that the spectral peaks are typically around a 
few hundred keVs. Therefore, for the purpose of comparing the energy fluxes between the prompt and flare, we estimate the prompt flux expected in the energy range 0.3 -10 keV by extrapolating the spectrum obtained from the 
analysis in the 15 -150 keV to those energies. In Figure \ref{fig:Energy_corr_multi}b, we have compared the peak fluxes of both prompt and X-ray flare estimated in the energy range for 0.3-10 keV. Again, no significant correlation between these parameters are found. The flares are found to be less brighter than the prompt emission, Figure \ref{fig:Energy_corr_multi}c. We further note that among the multiple flaring episodes of the GRB, the subsequent flaring episodes are found to be less intense as evident in Figure \ref{fig:Energy_corr_multi}d. This is consistent with the findings by \cite{chincarini2007} and \cite{Falcone_etal_2007}. \\
The prompt emission do not show any correlation between its peak flux and its duration (Figure \ref{fig:Energy_corr_multi}e). However, it is interesting to note that the flare peak flux shows a weak trend of negative correlation 
(Pearson correlation coefficient = -0.42) with flare duration (Figure \ref{fig:Energy_corr_multi}f). Similarly prompt emission does not show any significant correlation between it's peak flux and minimum variability and the quiescent period (Figure \ref{fig:Energy_corr_multi}g and \ref{fig:Energy_corr_multi}i). However, the peak flux of X-ray flare shows significant negative correlation between  its minimum variability and quiescent period (Figure \ref{fig:Energy_corr_multi}h and \ref{fig:Energy_corr_multi}j) where modelling is given as follows:

\begin{equation}
   log_{10}(F_{F,peak}) = (-0.89 \pm 0.05)\, log_{10}(T_{F,var}) - 7.44 \pm 0.06 
\end{equation}
\noindent

\begin{equation}
   log_{10}(F_{F,peak}) = (-1.05 \pm 0.07)\, log_{10}(T_{q}+1) -6.39 \pm 0.17 
\end{equation}
\noindent

Certain GRBs among the sample show the occurance of flaring partly or fully during the plateau phase of the afterglow emission (Appendix A). 
The plateau phase is considered to be the region where the power law slope of the afterglow ranges between +/- 0.3. We have identified 13 GRBs wherein 12 of them have either first or all episodes of flares occurring partly or fully during the plateau phase while only one case (GRB110414A) has its second episode of flaring alone, occuring during the plateau phase. 
These GRBs are particularly highlighted in all 
the correlation plots. We note that these GRBs do not exhibit any unusual trend in the correlation studies in comparison to the other GRBs in the sample.





\subsection{Clustering} \label{subsec:CLUSTER}
It is observed that among the GRB population only certain GRBs possess X-ray flaring episodes and among 
them certain GRBs are found to have multiple episodes of flaring, some with and without quiescent
periods etc. Such variability among the different GRBs suggest the possibility of the existence of 
subgroups among the GRBs with X-ray flaring.  

No major dissimilarity is observable between the distribution of the properties of the prompt emission of GRBs with and without flares, see Appendix \ref{Appendix3}. In addition, the 
Kolmogorov–Smirnov (KS) test shows that it is highly likely that GRBs with and without X-ray flares belong to the same population. The Principle component analysis done in section \ref{PCA1} shows that 
there is no explicit correlation between the studied prompt and flare properties. 
Therefore, we incorporated only the X-ray flare properties to investigate the possibility of clusters among the flaring GRBs. This included the following X-ray flare parameters: T$_{F,100}$, T$_{F,var}$, T$_{q}$, Fluence$_F$, F$_{F,Peak}$ and $\alpha_{F}$.



Unsupervised clustering was carried out using Gaussian Mixture Model (GMM), Hierarchical and K-means algorithms\footnote{The clustering packages available in scikit$-$learn \citep{scikit-learn} were used.}.
To account for the large difference between numerical values of each parameters in the dataset, 
the data was standardised to a fixed order of range. All the chosen parameters except 
$\alpha_F$ is converted to log scale and later the complete dataset is scaled and centered by 
subtracting each values by their mean and dividing them by their standard deviation. The standardised dataset was then fed into the different clustering algorithms. 
 
Based on the various evaluation metrics such as Silhouette score, Davies-Bouldin Index and Calinski-Harabasz Index, the K-means was found to be the best clustering algorithm for the dataset.    
The number of clusters in K-means were determined by two methods: Silhouette 
score \citep{Rousseeuw1987} and the Elbow method. 
The elbow was found to be at 4, signifying 4 possible subgroups within the GRBs with X-ray flares and the Silhouette 
score also peaked at 4 number of clusters.  
The obtained clusters are visualised in the principal component space as shown in Figure \ref{fig:cluster_PCA}a. For understanding the 
origin of the clusters, we have marked the cases that had $T_q =0$ in squares, GRBs with multiple flaring episodes with star 
markers and GRBs with flare in the plateau region of afterglow with yellow triangle marker. Since the $T_q$'s contribution in the PCA analysis is the largest in the fourth principal component, for the purpose of better visualising of the clusters, we chose to plot the clusters using PC4 versus PC1 in Figure \ref{fig:cluster_PCA}a.
The average values
of the parameters of each of the cluster is reported in the Table \ref{tab:Cluster Statistics}. The GRBs with flares during the plateau are found across the clusters and does not display any discernible peculiar behaviours.

\section{Discussions and Conclusions}
\label{result_discussion}
The origin of the late time X-ray flares is largely a mystery. Owing to the similar erratic behaviour of the lightcurves, the prompt emission and X-ray flares, they are believed to be powered by the central engine. Therefore, in this work using the extensive database of {\it Swift} observations of both prompt and X-ray flares via BAT and XRT instruments onboard respectively, we performed a systematic statistical study of the various temporal and spectral parameters characterising these two emissions. The sample consisted of 220 long GRBs. Among them 172 GRBs possessed single episodes of flare emission, whereas, 44 and 4 GRBs had double and triple flaring episodes respectively. Among all the GRBs in the sample, $96$ GRBs had their flaring episode observed synchronously in BAT as well.  
The temporal properties of the prompt and X-ray flare emissions were characterised by parameters such as total duration, minimum variability and the quiescent period between the episodes; and their spectral properties were characterised by parameters such as the low energy spectral power-law index, peak energy flux and energy fluences. 

\begin{figure}[ht]
\centering
\includegraphics[width=16cm,height=12cm,keepaspectratio]{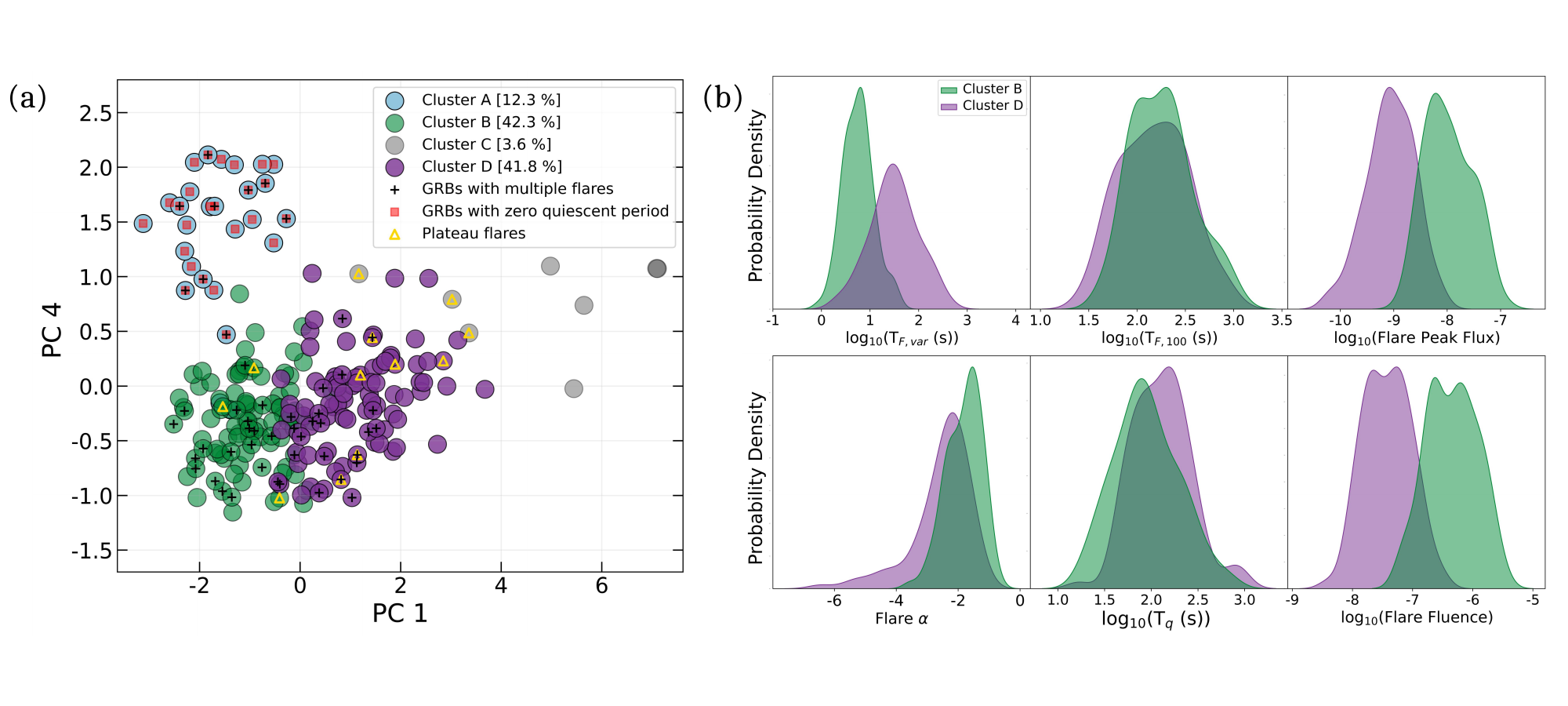}\\
\caption{(a) The K-means clustering results are visualized in a PC plot. 
Black$-$cross and red$-$square markers represent the GRBs which posses multiple flares and GRBs with zero quiescent period respectively. The yellow triangle markers represent GRBs with flares in the plateau region of afterglow. (b) The KDE distributions of the various flare properties of the cluster B and D are shown.}
\label{fig:cluster_PCA}
\end{figure}

\begin{table}[ht]
\caption{Cluster Statistics}
\label{tab:Cluster Statistics}
\begin{tabular}{lcccc}
\hline
              & Cluster A       & Cluster B        & Cluster C        & Cluster D      \\ \hline

log$_{10}(T_{F,100}$)         & 2.08 $\pm$ 0.46   &   2.26 $\pm$ 0.32    & 4.26 $\pm$  0.49  & 2.18 $\pm$ 0.35  \\
log$_{10}(T_{F,var}$)         & 0.97 $\pm$ 0.42   &   0.76 $\pm$ 0.31    & 3.14 $\pm$ 0.91   & 1.49 $\pm$ 0.49 \\
$\alpha_F$                    & -1.87 $\pm$ 0.59  &  -1.81 $\pm$ 0.56    & -1.64 $\pm$ 0.56  & -2.54 $\pm$ 0.95  \\
log$_{10}$(F$_{F,peak}$)      & -8.18 $\pm$ 0.43  &  -7.97 $\pm$ 0.44    & -10.39 $\pm$ 0.84 & -9.05 $\pm$ 0.44  \\
log$_{10}$(Fluence$_F$)       & -6.59 $\pm$ 0.49  &  -6.33 $\pm$ 0.41    & -6.70 $\pm$ 0.82  & -7.42 $\pm$ -0.39  \\
log$_{10}(T_{q}$)             & 0.07 $\pm$ 0.26   &  1.95 $\pm$ 0.35     & 3.54 $\pm$ 0.70   &  2.11 $\pm$  0.31   \\ \hline
Cluster size (\% of population)    & 12.3 \%         & 42.3 \%           & 3.6 \%            & 41.8 \%        \\ \hline
\end{tabular}
\end{table}

The distributions of the total duration of prompt and flaring episodes show that the flares largely have longer durations than the prompt with the median of $T_{F,100}$ distribution being nearly 3 times the median of $T_{P,100}$ distribution. The 
consecutive flaring episodes are found to be longer than the preceding one. The median of distribution of the duration of the second 
flaring episode is around 1.3 times the first episode of flare. This is consistent with the earlier studies done by \cite{chincarini2007}. 

The distribution of the minimum variability timescales of the prompt emission has a median around $2$ s. On the contrary, the X-ray flare emissions tend to have larger variability timescales with a median of 12 s. The quiescent period between the prompt emission and 
the flaring episodes are found to range from none to around a day. The average of quiescence for the first flaring episode is around 100 s. The spectral parameters such as the low energy power-law index in the flares is found to be much softer than the prompt 
emission. The study of the energy fluence and peak energy flux show that the prompt emission is more intense and brighter than the X-ray flaring episodes when compared in the same energy range of $0.3 - 10$ keV.  

The principal component analysis and the correlation studies carried out on the dataset reveals that there is no explicit correlation 
between the prompt and X-ray flare properties in general. The duration of the prompt and flare episodes are uncorrelated. 
Interestingly, a positive correlation is observed between the quiescent period and the flare duration, while there is no significant correlation with the prompt emission duration. 
This implies that the property which drives the quiescent period and the onset time of the flare emission after the prompt has ended, is intrinsically connected with the origin of the flare, while it is less dependent on the prompt emission.  

The minimum variability timescale observed in GRBs can point towards the size of the emitting region as well as allows one to understand the central engine activity, the impact of the circumburst medium as the jet 
propagates through the stellar core etc. 
There is no explicit correlation observed between 
the minimum timescale of variabilities of flares and that of prompt emission of the GRB. While there is no correlation between the prompt emission minimum variability and its duration, we find that 
there is a trend of positive correlation between the flare variability and its duration. 

We compared the spectral properties of 
prompt and flare in the energy range 0.3 - 10 keV. This was done by  extrapolating the spectrum obtained for prompt in the energy range 15-150 keV into the X-ray flare energy range of 0.3-10 keV. We find that the low energy power-law spectral index, $\alpha$ obtained for the spectrum of the total duration of prompt and flare respectively are found 
to be largely uncorrelated. The flare largely tends to have relatively much softer spectra in comparison to prompt. This may be attributed to the relatively longer duration of flares and may also point towards more 
rapidly evolving spectral behaviour in flares in comparison to prompt. We also find that the peak fluxes of prompt and flare emissions in the energy range 0.3-10 keV are uncorrelated. Consistent with previous 
studies \citep{Falcone_etal_2007}, the peak flux of the flare emission are found to be less brighter than the prompt and also decreases with consecutive flaring episodes. There is no explicit correlations 
observed between the peak fluxes of either of the prompt or flare and its durations. Interestingly, we also note that though there is no explicit correlation observed between prompt emission's minimum variability 
timescale and its peak flux, however, there exists a strong negative correlation between the flare's minimum variability timescale and its peak flux.

Further, using the flare properties alone, we explored the variety of classes among the current dataset of flares defined in this paper. Using K-means clustering algorithm, four clusters are 
found to exist. 
Among the clusters, we note that there is no preferential clustering for GRBs with multiple episodes of flares and 
are found to be spread across clusters. Among them we note that cluster A represents all $25$ cases of flares which have zero quiescent period (Figure \ref{fig:cluster_PCA}a). 
Thus, we conclude that the segregation of these X-ray flare events from the rest of the flare sample point out that these observed flaring episodes are actually the continuing prompt emission. 
Among the remaining classes, we note that cluster C contains only $3.6\%$ of the total sample and represents 
largely the flares with extreme values of duration, minimum variability timescale etc. This reduces the clustering 
to largely two main 
clusters B and D, each consisting $\sim 42\%$ of the sample.  While inspecting the distributions of the various properties of these two clusters, we find that 
the segregation is largely driven by minimum variability timescale, peak flux and fluence as evident from the Figure \ref{fig:cluster_PCA}b. 
Certain GRBs in the sample exhibit flaring during the plateau phase of the afterglow, but they do not display any unusual trends in correlation studies and are distributed across clusters without any discernible peculiar behaviours.

Thus, to summarise, in order to understand the origin of the prompt and X-ray flares, 
we find the following results of the correlation study:
(a) no explicit correlation between the temporal and the spectral properties of the prompt and flare emissions; 
(b) there exists a trend of positive correlation between minimum variability timescale of flare and its duration, while there is a strong negative correlation between the minimum variability timescale of flare and it peak flux. However, there are no such trends to be found in the prompt emission properties; and,  
(c) there also exists a positive correlation between quiescent period and flare duration, and a negative correlation between the quiescent period and the flare peak flux, however, no such 
correlations are observed for the prompt emission of GRBs.\\
\noindent
Therefore, using the current parameter characterisation of the prompt and X-ray flares adapted in this work, we do not find any significant evidence for the common origin for prompt emission of GRBs and the X-ray flares observed 
during the afterglow emission. In addition, we find the X-ray flares dominantly possess two main classes whose differences are primarily driven by the variability timescale, peak flux and 
fluences. Finally, we present the catalog of the estimated parameters to the community for further use in Appendix \ref{Catalogtable}.    

One of the major future prospects of this work is to develop an understanding of the physical scenario leading to the observed correlations between late-time X-ray flares and the prompt emission. Furthermore, for a better comprehensive picture of the origin of flares, an extensive comparative study need to be carried out between the flares observed in X-rays and optical as well as high energy emission ($>100$ MeV). 
The upcoming Cherenkov Telescope Array (CTA, \citealt{Kakuwa2012}) holds promise for further insights. We further note that a significant fraction of the initial sample had to be excluded due to incomplete data. An 
enhanced sample can be studied by adopting a light curve reconstruction method, similar to the one proposed in \cite{Dainotti_2023} as increased sample size can lead to more confident assessments of the correlations presented in this study.

\begin{acknowledgments}
We thank the anonymous referee for valuable comments and suggestions. J.S. and S.I. are supported by DST INSPIRE Faculty Scheme (IFA19-PH245). S.I. 
is also supported by SERB SRG Grant (SRG/2022/000211). This research has made use of data obtained from the High Energy Astrophysics Science 
Archive Research Center (HEASARC), provided by NASA's Goddard Space Flight Center and from the UK Swift Science Data Centre at the University of Leicester.
\end{acknowledgments}

\addtocontents{toc}{\protect\setcounter{tocdepth}{1}}
\newpage 
\appendix
\section{Additional Plots}
\label{Appendix4}

\begin{figure}[h!]
\centering
\includegraphics[width=\textwidth,keepaspectratio]{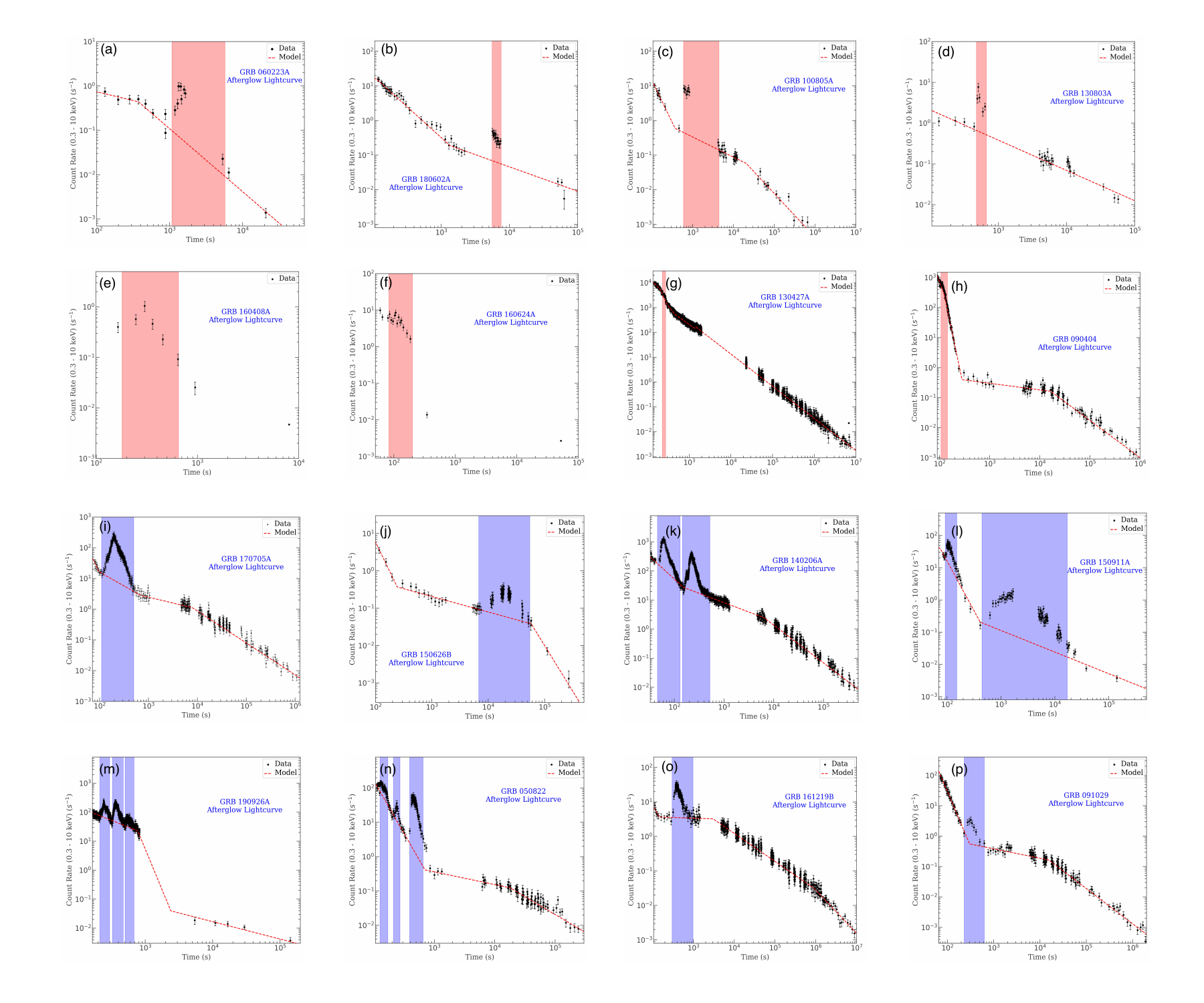}\\
\caption{The afterglow lightcurves of (a) GRB 060223A and (b) GRB 180602A represent cases with no data during the decay and rise phases of the flaring episode respectively.
(c) GRB 100805A and (d) GRB 130803A represent cases that have less observational points during flare emission.
(e) GRB 160408A and (f) GRB 160624A represent cases where the underlying afterglow emission is absent or not enough data points are available for proper modelling. 
(g) GRB 130427A and (h) GRB 090404 represent cases that are avoided due to their very low signal-to-noise ratio. The red shaded portion represents the flare duration as identified by the {\it Swift} XRT team. 
The afterglow lightcurves of (i) GRB 170705A and (j) GRB 150626B possessing a single flare are shown. (k) GRB 140206A and (l) GRB 150911A represents afterglows with double flaring episodes, while, (m) and (n) show lightcurves of GRB 190926A and GRB 050822, which have three flaring episodes each. 
(o) and (p) represent lightcurves of GRB 161219B and GRB 091029 with a full and partial flare, respectively, within the plateau region of their afterglow.
The blue-shaded region represents the flare duration as identified in this study. The red dotted line shown in all the plots represents the afterglow model obtained after excluding the flare region.   
}
\label{fig:lightcurves_combi}
\end{figure}

\section{Comparing the prompt emission of GRBs with and without X-ray flares}
\label{Appendix3}
\begin{figure}[h!]
\centering
\includegraphics[width=10cm,height=10cm,keepaspectratio]{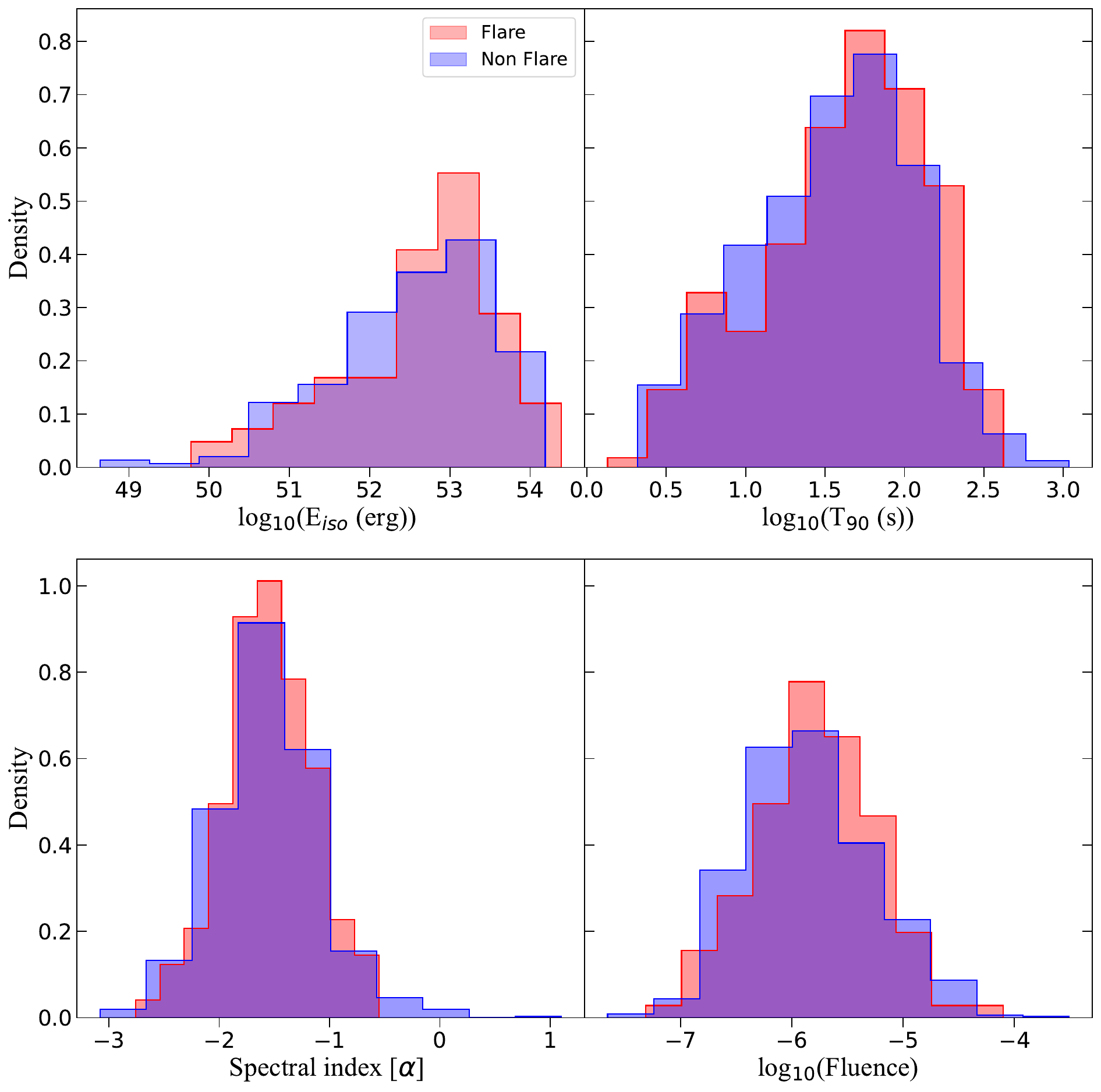}\\
\caption{Comparison of the distribution of the prompt emission properties such as $E_{iso}$, $T_{90}$, $\alpha$ and energy fluence of GRBs with and without X-ray flares are shown.}
\label{fig:prompt_comp}
\end{figure}

Not all GRBs detected by {\it Swift} possess X-ray flares in their afterglow emission. Thus, in order to understand if there 
exists any difference in the type of GRBs giving rise to flares, we made comparison 
plots of the distributions of 
different prompt emission properties such as isotropic burst energy, $E_{iso}$ estimated for known redshift cases, $T_{90}$, low energy power-law spectral index, $\alpha$ and energy fluence of GRBs with and without X-ray flares as shown in Figure \ref{fig:prompt_comp}. Furthermore, to assess if both the sample distributions originate from the 
same population, we conducted a Kolmogorov–Smirnov(KS) test\footnote{The scipy package ks$\_$2samp was used.} on the distributions of the different parameters. The null hypothesis assumes both the flare and non-flare GRB's 
properties are sampled from the same underlying population and the alternate is they are not identical. The resultant p-value if lower than the probability $p = 0.01$ (chosen threshold), 
allows us to reject the null hypothesis and adopt the alternate hypothesis at a statistical significance greater than $99\%$. However, for the distributions of $E_{iso}$, $T_{90}$, $\alpha$ and energy fluence in $15-150 \rm \,  keV$, we find the p-values from the KS test to be $0.17$, $0.07$, $0.69$ and $0.03$ respectively. Since the p-values lie above the threshold, we conclude that there is no particular distinction between the population of GRBs that produce and that does not produce late time X-ray flares. 


\clearpage

\clearpage

\section{Table of Prompt $\&$ X-ray Flare properties}
\label{Catalogtable}
The complete catalog of the studied parameters of the sample is made available online in the machine-readable format. 
A sample of the catalog is shown in Table \ref{catalog} for guidance regarding its form and content.


\begin{table}[h!]
\centering
\caption{catalog}
\label{catalog}
\resizebox{\linewidth}{!}{%
\begin{tabular}{lllllllllllllllllllll}
\hline
\multicolumn{1}{c}{GRB} & Flare & T$_{P,100}$ & T$_{P,Var}$ & T$_{P,Peak}$ & BFM [I] & \multicolumn{1}{c}{$\alpha_{P}$ [I]} & Fluence$_{P}$ & Fluence$_P$ & BFM [P]& \multicolumn{1}{c}{$\alpha_{P}$ [P]} & F$_{P,Peak}$ & F$_{P,Peak}$ & T$_{F,100}$ & T$_{F,Var}$ & T$_{F,Peak}$ & \multicolumn{1}{c}{$T_q$} & $S/N$ & \multicolumn{1}{c}{$\alpha_F$}  & F$_{F,Peak}$ &  Fluence$_F$ \\

 \multicolumn{1}{c}{NAME} & Episode  & (s) & (s) & (s) &  &  & [15-150keV] & [0.3-10keV]  &  &  & [0.3-10keV] & [15-150keV] & (s) & \multicolumn{1}{c}{(s)}  & \multicolumn{1}{c}{(s)}     & \multicolumn{1}{c}{(s)}  &  &  & [0.3-10keV] & [0.3-10keV] 
 \\ 
 &   &  &  &  &  &  & ($\times{10^{-6}}$$\rm erg/cm^2$) & ($\times{10^{-6}}$$\rm erg/cm^2$)  &  &  & ($\times{10^{-8}}$$\rm erg/cm^2/s$) & ($\times{10^{-8}}$$\rm erg/cm^2/s$) &  &  &      &  &  &  & ($\times{10^{-10}}$$\rm erg/cm^2$) & ($\times{10^{-8}}$$\rm erg/cm^2$) 
 \\ 
 \hline
170208B$^{\dag}$ & 1 & 32.4$\pm$2.4 & 0.31 & 1.0 & PL & -1.50$^{+0.01}_{-0.01}$ & 4.84$\pm$0.37 & 1.52$\pm$0.13 & PL & -1.08$^{+0.03}_{-0.02}$ & 6.76$^{+0.47}_{-0.44}$ & 74.27$^{+1.40}_{-1.39}$ & 172.2$\pm$31.4 & 6.37 & 98.1 & 70.5$\pm$3.37 & 28.9 & -2.88$^{+0.03}_{-0.03}$ & 338.86$\pm$44.40 & 46.55$\pm$8.53 \\

140419A & 1 & 124.6$\pm$9.3 & 0.36 & 51.4 & PL & -1.02$^{+0.10}_{-0.10}$ & 15.6$\pm$1.2 & 1.2$\pm$0.1 & PL & -1.02$^{+0.04}_{-0.04}$ & 3.24$^{+0.38}_{-0.32}$ & 42.89$^{+1.25}_{-1.30}$ & 106.3$\pm$9.9 & 10.84 & 209.9 & 60.1$\pm$10.6 & 6.0 & -1.42$^{+0.16}_{-0.16}$ & 22.45$\pm$5.00 & 4.50$\pm$0.96 \\

121125A & 1 & 78.2$\pm$7.9 & 1.07 & 36.6 & PL & -1.54$^{+0.03}_{-0.03}$ & 4.37$\pm$0.44 & 1.52$\pm$0.18 & PL & -1.37$^{+0.07}_{-0.07}$ & 4.35$^{+0.88}_{-0.73}$ & 21.43$^{+1.02}_{-1.08}$ & 44.5$\pm$7.4 & 18.9 & 92.0 & 15.7$\pm$8.0 & 5.3 & -1.76$^{+0.13}_{-0.14}$ & 22.07$\pm$4.42 & 2.73$\pm$0.56 \\


190926A$^{\dag}$ & 1 & 130.1$\pm$12.0 & 130.10 & 53.0 & PL & -2.06$^{+0.11}_{-0.10}$ & 1.46$\pm$0.17 & 2.74$\pm$0.89 & PL & -2.03$^{+0.40}_{-0.40}$ & 5.86$^{+15.57}_{-3.89}$ & 3.56$^{+0.78}_{-1.15}$ & 85.5$\pm$6.6 & 14.36 & 242.1 & 78.5$\pm$12.58 & 11.5 & -1.07$^{+0.07}_{-0.07}$ & 79.21$\pm$13.81 & 16.51$\pm$1.61 \\
190926A & 2 &  &  &  &  &  &  &  &  &  &    &  & 145.2$\pm$6.3 & 9.19 & 361.4 & 194.1$\pm$12.5 & 12.3 & -1.35$^{+0.04}_{-0.04}$ & 86.43$\pm$13.01 & 34.59$\pm$1.87 \\
190926A & 3 &  &  &  &  &  &  &  &  &   &  &  & 178.9$\pm$18.9 & 81.75 & 584.5 & 370.4$\pm$15.3 & 5.0 & -2.03$^{+0.06}_{-0.06}$ & 22.74$\pm$4.83 & 11.15$\pm$1.33 \\
121217A$^{\dag}$ & 1 & 101.4$\pm$16.7 & 3.51 & 0.0 & PL & -1.54$^{+0.04}_{-0.05}$ & 2.01$\pm$0.34 & 0.71$\pm$0.15 & PL & -1.30$^{+0.12}_{-0.11}$ & 1.75$^{+0.65}_{-0.46}$ & 10.39$^{+0.77}_{-0.83}$ & 764.3$\pm$2.2 & 6.34 & 737.7 & 237.5$\pm$16.83 & 18.3 & -1.41$^{+0.01}_{-0.01}$ & 146.36$\pm$18.90 & 189.24$\pm$2.15 \\
080906 & 1 & 180.0$\pm$12.4 & 4.37 & 11.0 & PL & -1.61$^{+0.04}_{-0.04}$ & 3.65$\pm$0.27 & 1.61$\pm$0.22 & PL & -1.60$^{+0.15}_{-0.15}$ & 3.61$^{+1.87}_{-1.21}$ & 8.29$^{+0.80}_{-0.89}$ & 103.2$\pm$14.1 & 45.02 & 186.4 & 60.2$\pm$12.95 & 6.0 & -2.09$^{+0.05}_{-0.06}$ & 21.40$\pm$3.70 & 6.52$\pm$0.94 \\
080906 & 2 &  &  &  &  &  &  &  &  &  &     &  & 125.1$\pm$34.9 & 115.54 & 596.2 & 447.6$\pm$24.4 & 1.8 & -3.30$^{+0.41}_{-0.41}$ & 2.30$\pm$0.58 & 0.33$\pm$0.17 \\
200509A$^{\dag}$ & 1 & $40.0\pm3.5$ & 0.07 & 1.8 & PL & -1.08$^{+0.10}_{-0.10}$ & $1.46\pm0.15$ & $0.13\pm0.04$ & PL & -0.58$^{+0.27}_{-0.28}$ & 0.25$^{+0.31}_{-0.13}$ & 11.65$^{+1.37}_{-1.67}$ & $810.2\pm20.1$ & 2.65 & 824.5 & $220.1\pm18.35$ & 26.1 & -1.80$^{+0.01}_{-0.01}$ & $293.63\pm35.30$ & $231.45\pm5.98$ 
\\
\hline
\multicolumn{17}{l}{Notes. BFM - Best Fit Mode; [I] - Integrated; [P] - Peak; $S/N$ - signal to noise ratio; \dag - GRBs that had their flaring episodes observed synchronously in BAT }
\end{tabular}%
}
\end{table}

\bibliography{References}{}
\bibliographystyle{aasjournal}



\end{document}